\documentclass[nofootinbib, showkeys, preprint] {revtex4}

\usepackage{epsf, epsfig, subfigure, graphicx, amsmath, amssymb}
\usepackage{color}

\begin{document}


\title{Assisted freeze-out}

\author{Genevi\`eve B\'elanger$^1$ and Jong-Chul Park$^{1, 2}$}

\address{$^1$LAPTH, U. de Savoie, CNRS, BP 110, 74941 Annecy-Le-Vieux, France\\
$^2$Korea Institute for Advanced Study, Seoul 130-722, Korea}

\begin{abstract}
We explore a class of dark matter models with two dark matter
candidates, only one interacts with the standard model sector. One
of the dark matter is thermalized with the assistance of the other
stable particle. While both stable particles contribute to the
total relic density only one can elastically scatter with nuclei,
thus effectively reducing the direct detection rate.

\end{abstract}


\maketitle


\section{Introduction}

The most natural explanation for the astrophysical and
cosmological indications of a large component of dark matter in
the Universe is a new weakly interacting massive particle that
behaves as cold dark matter. Such dark matter candidates are found
in different  extensions of the standard model
(SM)~\cite{Bertone:2004pz}. However, dark matter (DM) could be
composed of more than one particle, and several multi component DM
models have been suggested recently~\cite{Zurek:2008qg,
Profumo:2009tb, Feldman:2010wy, Winslow:2010nk, Batell:2010bp,
Liu:2011aa, Adulpravitchai:2011ei}. Interest for multi component
DM arose, in particular, from hints for signals in cosmic rays as
well as in direct detection experiments corresponding to two
completely different DM mass scales. On the one hand, the PAMELA
experiment~\cite{PAMELA} has reported an cosmic ray excess of
positrons which would be consistent with a TeV scale DM candidate
annihilating mainly into leptons. On the other hand,
DAMA~\cite{DAMA}, CoGeNT~\cite{Cogent} and CRESST~\cite{CRESST}
have all reported excesses in the direct detection rate that would
be compatible with DM around 10 GeV. Even though there is no
conclusive evidence that these signals are due to DM, it is
interesting to investigate multi component DM models to expand the
range of DM models at a time when DM searches in indirect, direct
and collider experiments are increasing their sensitivities.
Furthermore, replacing the usual R-parity symmetry that guarantees
the stability of the lightest R-parity odd particle by an enlarged
symmetry group allows not only multiple DM candidates but also new
freeze-out mechanisms. In particular, the semi-annihilation
mechanism where two DM particles annihilate into another DM
particle and a SM particle was proposed in
Ref.~\cite{semiannihilation}.


In this paper, we propose a new type of freeze-out mechanism,
assisted freeze-out, within the framework of multi component DM
models. In this new freeze-out mechanism, one DM candidate can be
thermalized only through the assistance of the other stable
particle. Thus, the decoupling of one DM particle from the thermal
bath is influenced by the other DM particle. Consequently, the
relic density of DM is solved by using two coupled Boltzmann
equations for two stable fields. In the analysis of the
right-handed sneutrino DM~\cite{arXiv:1105.1652}, a similar
situation has been already considered even though this model has
only one stable particle. After setting the Boltzmann equations,
we construct a simple model with two hidden DM sectors
corresponding to two new $U(1)$ gauge symmetries, only one of
which interacts with the SM sector. This is achieved through
kinetic mixing of the new and standard gauge bosons. The DM
particles are assumed to be Dirac fermions. We then show how both
particles can contribute to the relic density of DM while only one
can scatter elastically on nucleons. Although the direct detection
rate tends to be rather high in this model, we show examples,
where all constraints can be satisfied, including a case with a DM
candidate around 10 GeV.

This paper is organized as follows. The basic set-up is presented
in section 2. An explicit model is constructed in section 3 and
the implications for the relic density of DM as well as for the
direct detection rates on nucleons are studied. Section 4 contains
our conclusions.

\section{Basic set-up}

We consider the case where $\chi_1$ and $\chi_2$ are stable dark
matter candidate particles. This can be achieved, for example,
with a $Z_2 \otimes Z'_2$ symmetry. We assume that $m_2 > m_1$
with $m_i = m_{\chi_i}$ and that $\chi_2$ can only annihilate into
$\chi_1$ and not into SM particles.

A set of coupled Boltzmann equations describe the evolution of the
number density $n_i$ of particle $\chi_i$. In the following, $X$
stands for some SM particles and  $m_X < m_1$.
\begin{eqnarray}
\frac{dn_2}{dt} + 3Hn_2 &=&
- \langle \sigma v \rangle_{22 \rightarrow 11}
\left[ (n_2)^2 - \frac{(n_2^{\rm eq})^2}{(n_1^{\rm eq})^2} (n_1)^2 \right]\;,
\label{Boltzmann1}\\
\frac{dn_1}{dt} + 3Hn_1 &=&
- \langle \sigma v \rangle_{11 \rightarrow XX}
\left[ (n_1)^2 - (n_1^{\rm eq})^2 \right]
- \langle \sigma v \rangle_{11 \rightarrow 22}
\left[ (n_1)^2 - \frac{(n_1^{\rm eq})^2}{(n_2^{\rm eq})^2} (n_2)^2 \right] \nonumber\\
&=& - \langle \sigma v \rangle_{11 \rightarrow XX}
\left[ (n_1)^2 - (n_1^{\rm eq})^2 \right]
+ \langle \sigma v \rangle_{22 \rightarrow 11}
\left[ (n_2)^2 - \frac{(n_2^{\rm eq})^2}{(n_1^{\rm eq})^2} (n_1)^2 \right]\;,
\label{Boltzmann2}
\end{eqnarray}
where $H$ is the Hubble parameter and $n^{\rm eq}_i$ is the
equilibrium number density of particle $i$. In solving the
Boltzmann equations (\ref{Boltzmann1}) and (\ref{Boltzmann2}), it
is useful to introduce  the variable $Y_i \equiv n_i/s$ describing
the actual number of particle $i$ per comoving volume, where $s$
is the entropy density of the Universe. Solving these coupled
Boltzmann equations, one can find $Y_i$ as a function of $x \equiv
m_1/T$.\footnote{Note that as opposed to the familiar case of
coannihilation~\cite{coannihilation}, one cannot assume that
$n_1^{eq}/n_2^{eq} \simeq n_1/n_2$ which relies on the fact that
scattering with background SM particles, e.g. $\chi_2 X
\rightarrow \chi_1 X'$, occurs at a high rate. In the assisted
freeze-out scenario, such reactions do not take place. Thus, we do
not use this approximation.} In the new variables, the Boltzmann
equations are recast as
\begin{eqnarray}
\frac{dY_2}{dx} &=&
- x^{-2} \lambda_{22 \rightarrow 11}
\left[ (Y_2)^2 - \frac{(Y_2^{\rm eq})^2}{(Y_1^{\rm eq})^2} (Y_1)^2 \right]\;,
\label{NewBoltzmann1}\\
\frac{dY_1}{dx} &=&
- x^{-2} \lambda_{11 \rightarrow XX}
\left[ (Y_1)^2 - (Y_1^{\rm eq})^2 \right]
+ x^{-2} \lambda_{22 \rightarrow 11}
\left[ (Y_2)^2 - \frac{(Y_2^{\rm eq})^2}{(Y_1^{\rm eq})^2} (Y_1)^2 \right]\;,
\label{NewBoltzmann2}
\end{eqnarray}
where
\begin{equation}
\lambda_{ij \rightarrow kl} \equiv \left[ \frac{s}{H} \right]_{x=1}
\langle \sigma v \rangle_{ij \rightarrow kl}(x)\;.
\label{lambda}
\end{equation}
The equilibrium number of particle $i$ per comoving volume
$Y_i^{\rm eq} \equiv n_i^{\rm eq}/s$ has the following forms:
\begin{eqnarray}
Y_1^{\rm eq} = \frac{g_1}{g_{*s}} \frac{45}{4\pi^4} x^2 K_2[x]\;,\quad
Y_2^{\rm eq} = \frac{g_2}{g_{*s}} \frac{45}{4\pi^4} (rx)^2 K_2[rx]\;,
\label{Yeq}
\end{eqnarray}
where $r \equiv m_2/m_1$, $g_i$ is the number of internal degrees of freedom of particle
$i$ and $K_2[x]$ is the modified Bessel function.

The heavier particle $\chi_2$ will not be in thermal equilibrium
during freeze-out unless $\langle \sigma v \rangle_{22 \rightarrow
11} \ll \langle \sigma v \rangle_{11 \rightarrow XX}$. Thus,
solving the Boltzmann equation in the case where many different
processes contribute to annihilation would be complicated. In this
analysis, we assume that $s-$wave annihilation processes dominate.
This allows to simplify the computation of the thermally averaged
cross section for $\chi_2$ when solving the Boltzmann equation
while illustrating the dependence of the relic density on each
parameter.

\begin{figure}
\begin{center}
\includegraphics[width=0.49\linewidth]{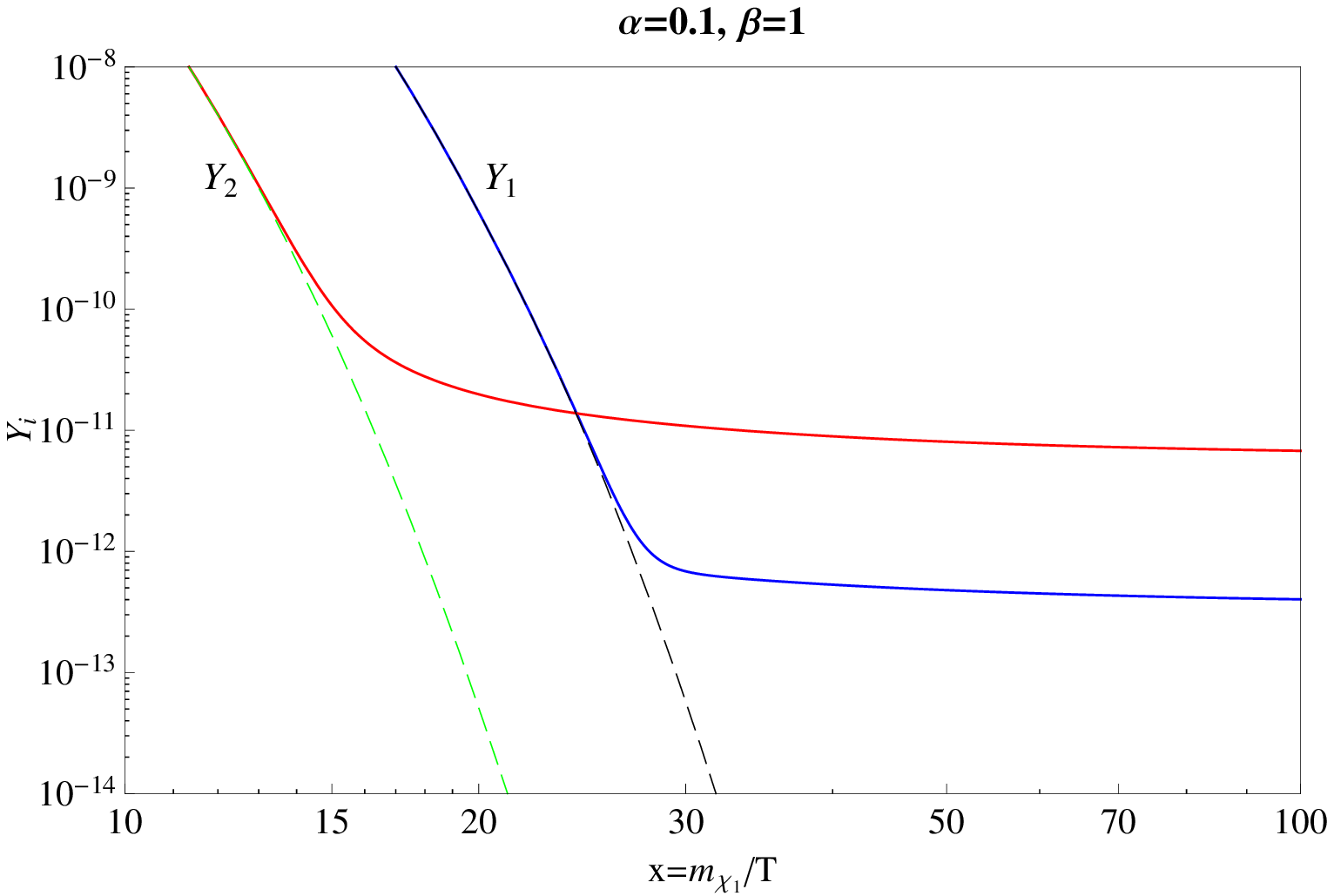}
\includegraphics[width=0.49\linewidth]{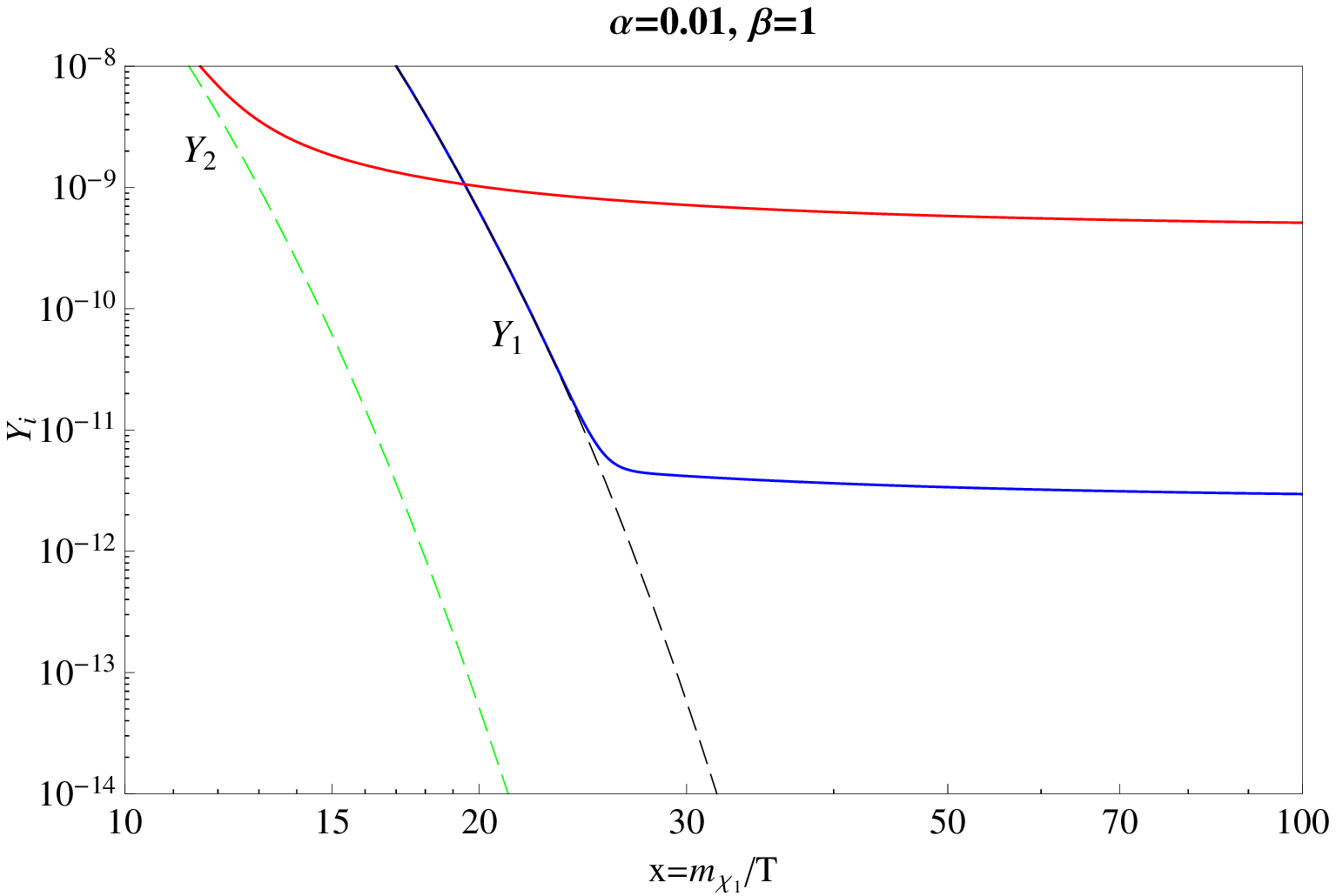}

\includegraphics[width=0.49\linewidth]{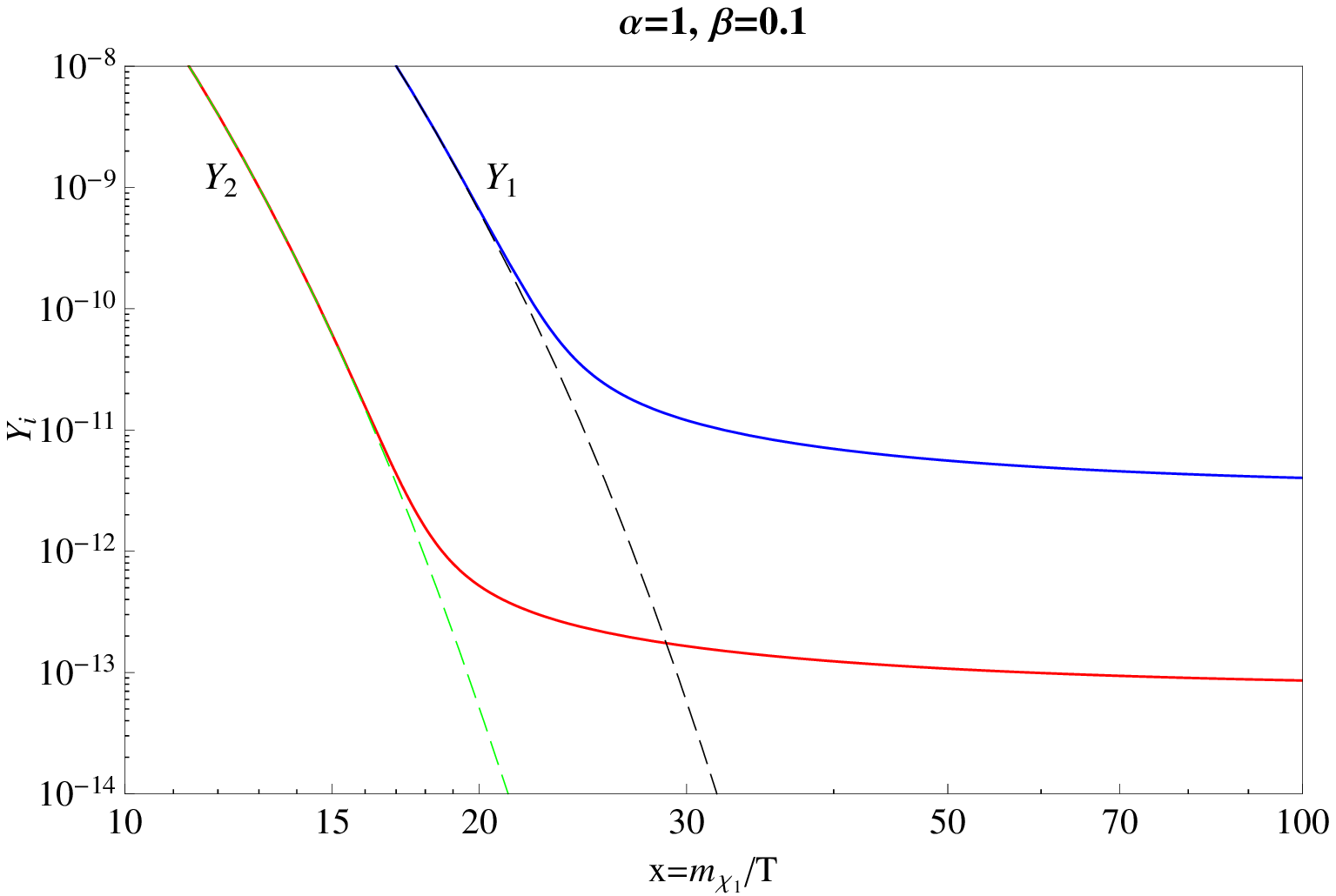}
\end{center}
\caption{The evolution of the abundances of $\chi_1$ (blue) and
$\chi_2$ (red)  per comoving volume as a function of $x \equiv
m_1/T$. Dashed lines are the equilibrium comoving number
densities. Dark matter masses are fixed as $m_1=100$ GeV and
$m_2=150$ GeV. The corresponding matrix elements are shown in each
figure. } \label{Fig1}
\end{figure}

If we limit our analysis to $s$-wave annihilation for simplicity,
we can simply express the relevant $s$-wave matrix elements as
\begin{equation}
\alpha \equiv \mathcal{M}_{22 \rightarrow 11} = \mathcal{M}_{11 \rightarrow 22}\;,
\quad \beta \equiv \mathcal{M}_{11 \rightarrow XX}\;.
\label{MatrixElements}
\end{equation}
It is then straightforward to solve the two coupled Boltzmann
equations (\ref{NewBoltzmann1}) and (\ref{NewBoltzmann2}) to
obtain the abundance of $\chi_1$ and $\chi_2$. To illustrate the
assisted freeze-out mechanism, we choose the DM masses $m_1=100$
GeV and $m_2=150$ GeV, taking  different values for the
annihilation amplitudes $\alpha$ and $\beta$. The evolution of the
abundances $Y_1, Y_2$ are displayed in fig.~\ref{Fig1} for
$\alpha$ and $\beta$: $(\alpha, \beta) = (0.1, 1)$, $(0.01, 1)$
and $(1, 0.1)$.
%
%
%
%
The blue and red solid lines are respectively $Y_1$ and $Y_2$, and
dashed lines show the corresponding equilibrium comoving number
densities. When $\chi_2$ interacts weakly with $\chi_1$ (see the
top panels of fig.~\ref{Fig1}), the abundance of $\chi_1$ is
mainly determined by its interaction with SM particles and due to
the weak interactions the final DM abundance is dominated with
$\chi_2$. A comparison of  the cases $\alpha=0.01$ and
$\alpha=0.1$ shows that the abundances of both particles are reduced
by the interactions between $\chi_1$ and $\chi_2$. When the
interactions of $\chi_1$ with  $\chi_2$ are stronger than with SM
particles, i.e. $\alpha \gg \beta$ (the bottom panel of
fig.~\ref{Fig1}), the freeze-out of $\chi_2$ is delayed and its
abundance is, consequently, much reduced. Because of the
interactions with $\chi_2$, the abundance of $\chi_1$ increases in
comparison with the previous case. The abundance of $\chi_1$ now
largely dominates over that of $\chi_2$.

\section{Dark matter candidates with two extra $U(1)$'s}

As an explicit example of how the assisted freeze-out mechanism
can occur in a specific DM model, we consider a toy model with  a
hidden sector containing two extra Abelian gauge symmetries,
$U(1)'$ and $U(1)''$, and two Dirac fermions $\psi_1$ and
$\psi_2$. The particle $\psi_1$ is charged under both $U(1)'$ and
$U(1)''$ gauge symmetries, and $\psi_2$ is only charged under the
$U(1)''$ symmetry. If $\psi_1$ is  the lightest particle charged
under the $U(1)'$ symmetry and $\psi_2$ is the lightest one only
charged under the $U(1)''$ symmetry, then $\psi_1$ and $\psi_2$
can be naturally stable particles like the electron in the SM. We
assume that the hidden sector couples to the SM sector only
through a kinetic mixing between $U(1)'$ and $U(1)_Y$. Then, the
full Lagrangian including this kinetic mixing is
\begin{eqnarray} \label{toyLagrangian}
{\cal L} = {\cal L}_{SM} &-& {1\over2} \sin\epsilon\, \hat{B}_{\mu\nu} \hat{X}'^{\mu\nu} -\frac{1}{4}\hat{X}'_{\mu\nu}\hat{X}'^{\mu\nu} -\frac{1}{4}\hat{X}''_{\mu\nu}\hat{X}''^{\mu\nu}
+ {1\over2} m_{\hat{X}'}^2 \hat{X}'^2 + {1\over2} m_{\hat{X}''}^2 \hat{X}''^2 \nonumber\\
&-& g_{X'} \hat{X}'_\mu \bar{\psi_1}\gamma^\mu \psi_1 - g_{X''} \hat{X}''_\mu \bar{\psi_1}\gamma^\mu \psi_1 - g_{X''} \hat{X}''_\mu \bar{\psi_2}\gamma^\mu \psi_2
+ m_1 \bar{\psi_1}\psi_1 + m_2 \bar{\psi_2}\psi_2\,,\quad\,\,
\end{eqnarray}
where the hidden $U(1)$'s are assumed to be spontaneously broken
leading to the gauge boson masses $m_{\hat{X}'}$ and
$m_{\hat{X}''}$. In this toy model, we will neglect the possible
effect of a mixing between the SM and hidden sector Higgs fields
for the simplicity of analysis. In the SM sector, the mass of the
$\hat{Z}$ gauge boson is $m_{\hat{Z}}$ and the gauge couplings are
denoted by $\hat{g}=\hat{e}/s_{\hat{W}}$ and $\hat{g}'=
\hat{e}/c_{\hat{W}}$.

The kinetic and mass mixing terms are diagonalized away by the following transformation:
\begin{eqnarray} \label{transformation}
\hat{B} &=& c_{\hat{W}} A - (t_\epsilon s_\xi+ s_{\hat{W}} c_\xi) Z
+ (s_{\hat{W}} s_\xi-t_\epsilon c_\xi) Z' \,, \nonumber\\
\hat{W}_3 &=& s_{\hat{W}} A + c_{\hat{W}} c_\xi Z
- c_{\hat{W}} s_\xi Z' \,, \nonumber\\
\hat{X}' &=& {s_\xi \over c_\epsilon} Z + {c_\xi\over c_\epsilon} Z' \,, \nonumber\\
\hat{X}'' &=& Z'' \,,
\end{eqnarray}
where the rotation angle $\xi$ is determined by
\begin{equation} \label{t_2csi}
\tan2\xi = - {m_{\hat{Z}}^2 s_{\hat{W}} \sin2\epsilon \over
               m_{\hat{X}'}^2 - m_{\hat{Z}}^2
               (c^2_\epsilon-s^2_\epsilon s_{\hat{W}}^2) }
\end{equation}
and the weak mixing angle $s_{\hat{W}}$ is very close to the
physical value $s_W$ due to the stringent $\rho$ parameter
constraint. Then, the $Z$ and $Z'$ gauge bosons obtain the
redefined masses:
\begin{eqnarray}
m_Z^2 &=& m_{\hat{Z}}^2(1+s_{\hat{W}} t_\xi t_\epsilon) \,,  \label{eq:mzhat} \\
m_{Z'}^2 &=&
   { m_{\hat{X}'}^2 \over c_\epsilon^2 (1+s_{\hat{W}} t_\xi t_\epsilon)} \,;
\label{eq:mx}
\end{eqnarray}
on the other hand, the $Z''$ gauge boson mass is the same as the
$\hat{X}''$ gauge boson: $m_{Z''}^2 = m_{\hat{X}''}^2$.

Let us list all the interaction vertices of the $W, Z, Z'$ and
$Z''$ gauge bosons relevant for our analysis. The $Z''$ boson has
the following non-modified couplings:
\begin{equation} \label{Z2fermion}
{\cal L} = - g_{X''} Z''_\mu [\bar{\psi_1}\gamma^\mu \psi_1 + \bar{\psi_2}\gamma^\mu \psi_2]\,.
\end{equation}
However, the other gauge bosons have modified couplings. In order
to describe the interaction vertices of $W, Z$ and $Z'$, let us
define the various couplings, $g$'s, as follows:
\begin{eqnarray}\label{g's}
 {\cal L} &=& W_\mu^+\, g_f^W [ \bar{\nu} \gamma^\mu P_L e
                        + \bar{u} \gamma^\mu P_L d ] + c.c. \nonumber \\
  &+& Z_\mu \left[ g^Z_{fL}\, \bar{f} \gamma^\mu P_L f
                          + g^Z_{fR}\, \bar{f} \gamma^\mu P_R f
       + g^Z_{\psi_1}\, \bar{\psi_1}\gamma^\mu\psi_1 \right] + g_W^Z [[Z W^+ W^-]] \nonumber \\
  &+& Z'_\mu \left[ g^{Z'}_{fL}\, \bar{f} \gamma^\mu P_L f
                + g^{Z'}_{fR}\, \bar{f} \gamma^\mu P_R f
       + g^{Z'}_{\psi_1}\, \bar{\psi_1}\gamma^\mu\psi_1 \right] + g_W^{Z'} [[Z' W^+ W^-]] \nonumber \\
  &+& h \left[  g^h_{ZZ}\, Z_\mu Z^\mu + g^h_{Z'Z'} Z'_\mu Z'^\mu + g^h_{Z'Z} Z'_\mu Z^\mu
       \right] \,.
\end{eqnarray}
One can find these redefined couplings expressed by the physical
observables (unhatted parameters) in the appendix of
\cite{Chun:2010ve} which describes a hidden sector with only a
$U(1)'$ symmetry. The only difference between the gauge boson
sectors of two models is the additional gauge boson associated
with $U(1)''$ that we assume to be decoupled from the SM sector.

\subsection{Relic density}

The relic abundances of two stable particles $\psi_{1, 2}$ are
determined by solving two coupled Boltzmann equations
(\ref{NewBoltzmann1}) and (\ref{NewBoltzmann2}). The thermal relic
density of DM is the sum of the relic densities of the two
candidates $\psi_{1, 2}$, $\Omega_{\rm DM} h^2= \Omega_{\psi_1}h^2
+ \Omega_{\psi_2}h^2$. The free parameters of the model  include
the mass parameters $m_1, m_2, m_{Z'}$ and $m_{Z''}$ as well as
the  hidden gauge couplings $g_{X'}$ and $g_{X''}$ and the kinetic
mixing parameter $\sin \epsilon$. In order to examine the
dependence of the relic abundance of $\psi_{1, 2}$ on each
parameter, in our numerical analysis we search for the parameter
space which satisfies the observed DM relic density
limit~\cite{WMAP7}, in the $m_1 - m_{Z''}$ plane fixing the other
parameters.\footnote{In all the numerical analysis, we assume that
$m_{Z''} > 2 m_1$ to avoid a stable $Z''$. Moreover, for the $Z'$
boson, electroweak precision test bounds and LHC detection
prospects were already studied in Ref.~\cite{Chun:2010ve, Mixing}.
We fix the kinetic mixing to its experimental upper bound given
in~\cite{Chun:2010ve, Mixing}. Smaller values of
$g_{X'}\sin\epsilon$ are, of course, even less constrained.}

The annihilation of $\psi_1$ pairs into SM particles proceeds
through $s-$channel exchange of $Z$ and $Z'$: possible final
states are $f\bar{f}, W^+W^-$ and $Zh$. The dominant annihilation
mode depends on the DM mass $m_1$, into only fermions when $m_1 <
M_W$ into additionally gauge bosons otherwise. In addition, all
amplitudes are proportional to the kinetic mixing $\sin\epsilon$
and the coupling $g_{X'}$. The $s-$channel exchange can be
strongly enhanced by a resonance effect. We therefore expect the
relic density to drop rapidly when $m_1 \approx m_Z/2$ or
$m_{Z'}/2$. The annihilation of $\chi_2$ proceeds uniquely through
the exchange of $Z''$ and is therefore determined solely by the
coupling $g_{X''}$ and the masses $m_{Z''}, m_2$.

\begin{figure}
\begin{center}
\includegraphics[width=0.47\linewidth]{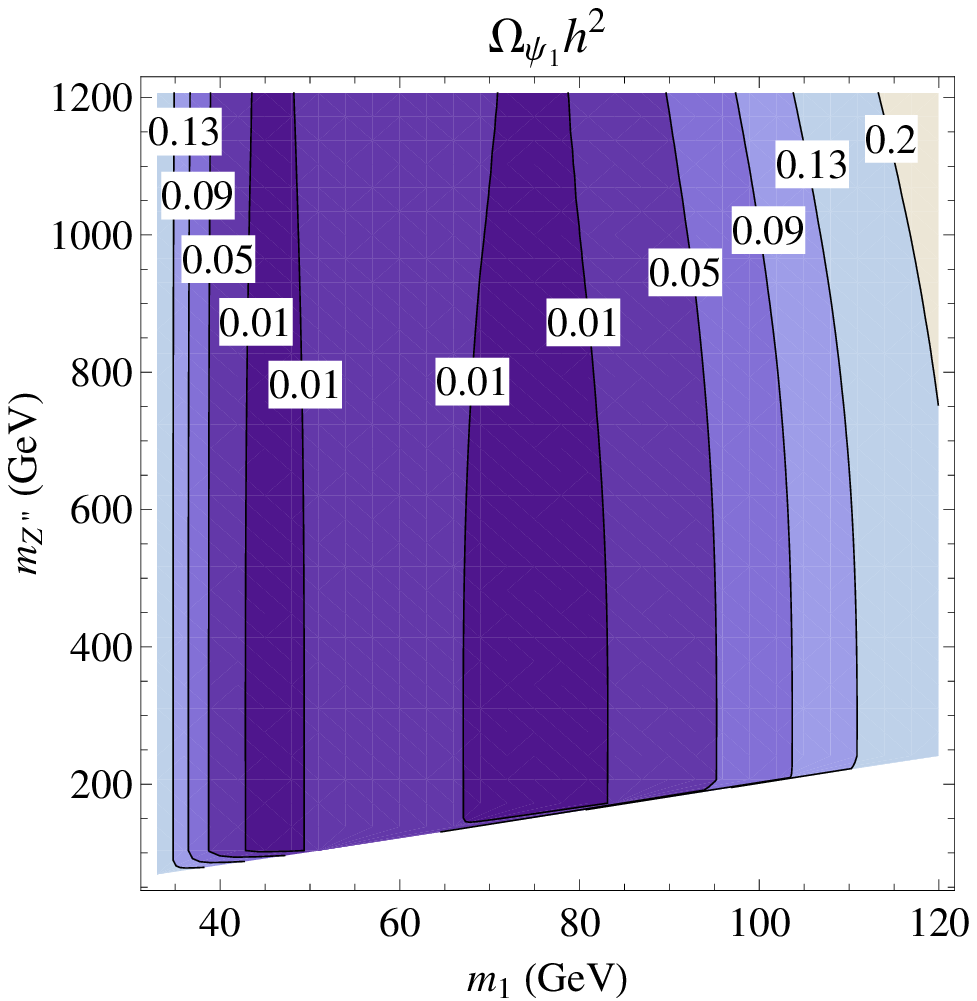}
\includegraphics[width=0.47\linewidth]{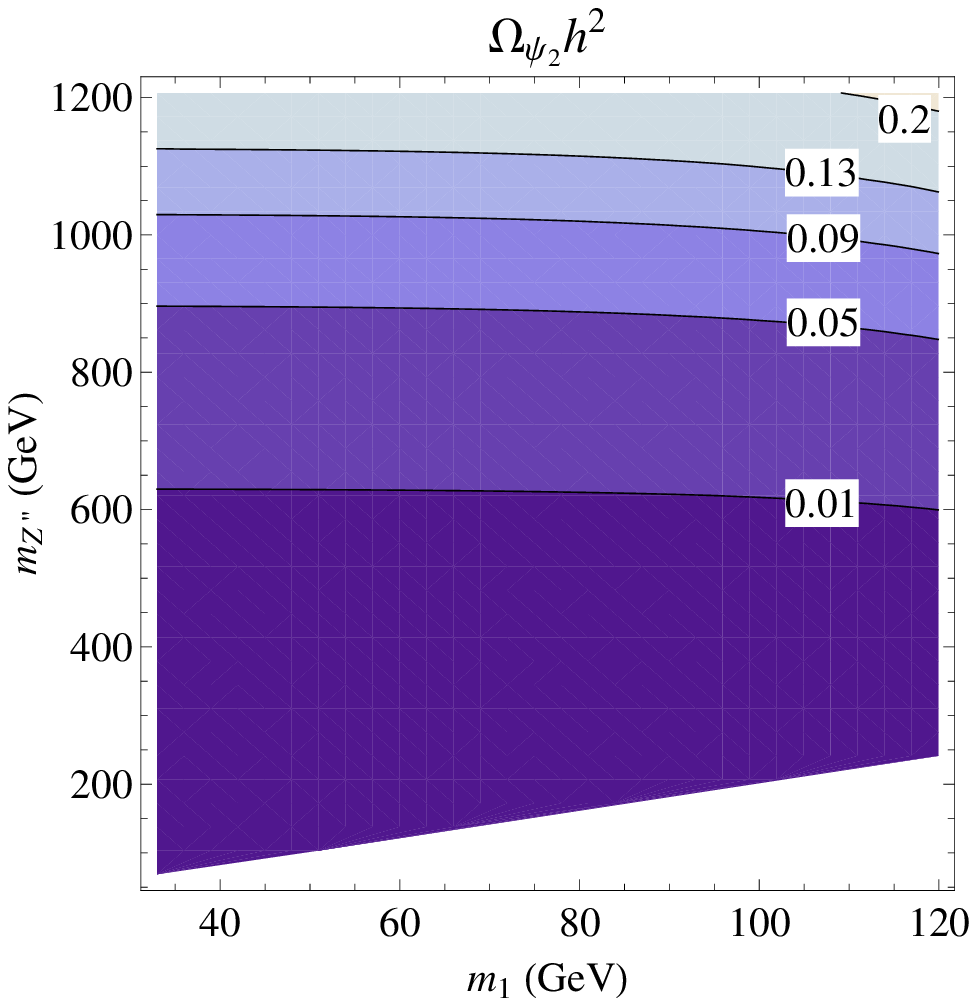}

\includegraphics[width=0.47\linewidth]{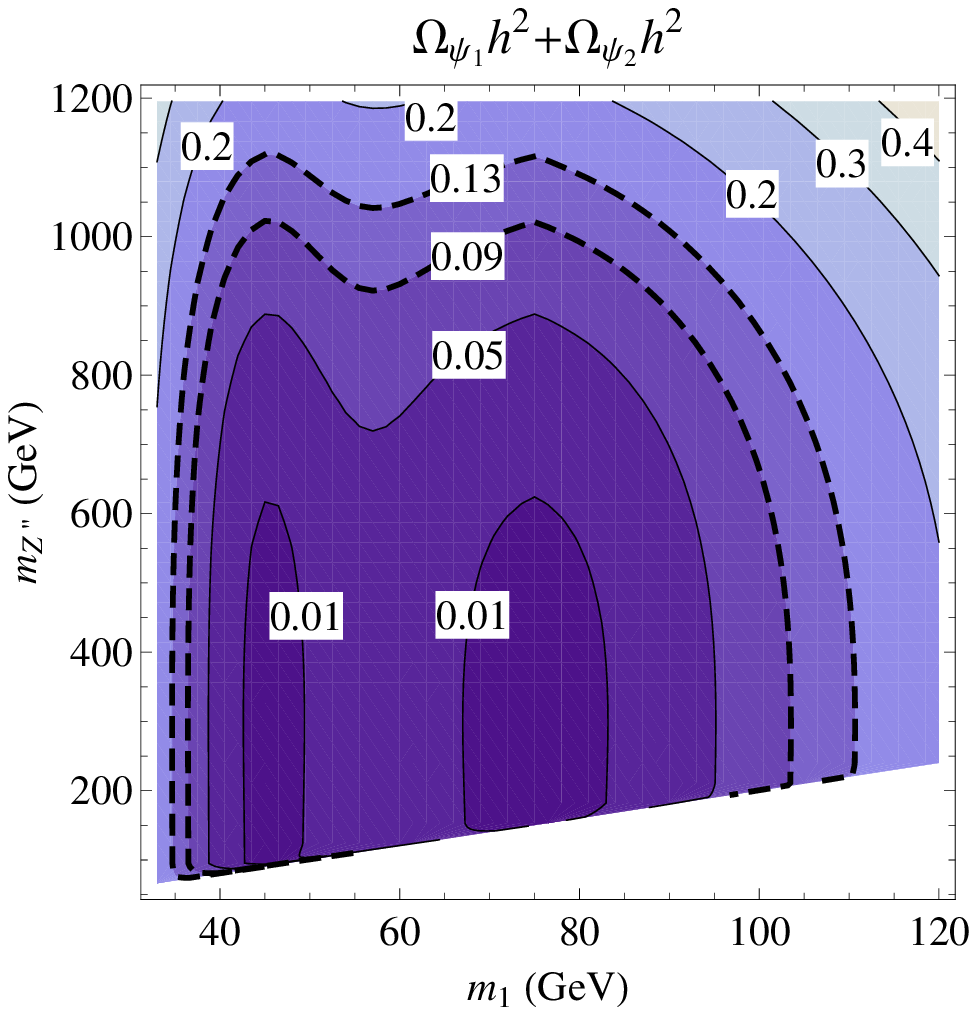}
\end{center}
\caption{Contour plots for the relic abundances of the dark matter
particles $\psi_1$ (top-left), $\psi_2$ (top-right) and the total
(bottom) in the $m_1 - m_{Z''}$ plane. We fix the other parameters
as follows: $g_{X'} = 0.5$, $g_{X''} = 0.9$, $m_2 = 150$ GeV and
$m_{Z'} = 150$ GeV. In the bottom panel, the region between two
thick dashed lines is allowed by the WMAP result on the DM relic
density.}\label{fig:reference}
\end{figure}

The relic densities $\Omega_{\psi_1}h^2$ and $\Omega_{\psi_2}h^2$
are displayed in fig.~\ref{fig:reference}, in the $m_1 - m_{Z''}$
plane for the reference case of $g_{X'} = 0.5, g_{X''} = 0.9, m_2
= 150$ GeV and $m_{Z'} = 150$ GeV. The enhanced annihilation near
the $Z~(Z')$ resonances explains the drop in $\Omega_{\psi_1}h^2$
for $m_1 \approx 45~(75)$~GeV. Similarly, the enhanced
annihilation near the $Z''$ resonance means that
$\Omega_{\psi_2}h^2$ is small for $m_{Z''} \approx 300$~GeV. Note
that there is a lower bound, $m_{Z''} > 2 m_1$ because we impose
the condition that the $Z''$ is not stable. The total DM relic
density $\Omega_{\psi_1}h^2 + \Omega_{\psi_2}h^2$ is displayed in
the bottom panel of fig.~\ref{fig:reference}. The region between
two thick dashed lines represents the points consistent with the
recent WMAP relic density result~\cite{WMAP7}, allowing for
$3\sigma$ experimental uncertainties. In two bands at low values
of $m_{Z''}$ where $m_1 \approx 35$~GeV and $m_1\approx 110$~GeV,
the DM relic density is dominated by $\psi_1$ while
$\Omega_{\psi_2} h^2$ dominates for $m_{Z''} \approx 1$~TeV. In
the transition between these regions, both particles give a
significant contribution to the total DM relic density.

%
%
%
%

\begin{figure}
\begin{center}
\includegraphics[width=0.47\linewidth]{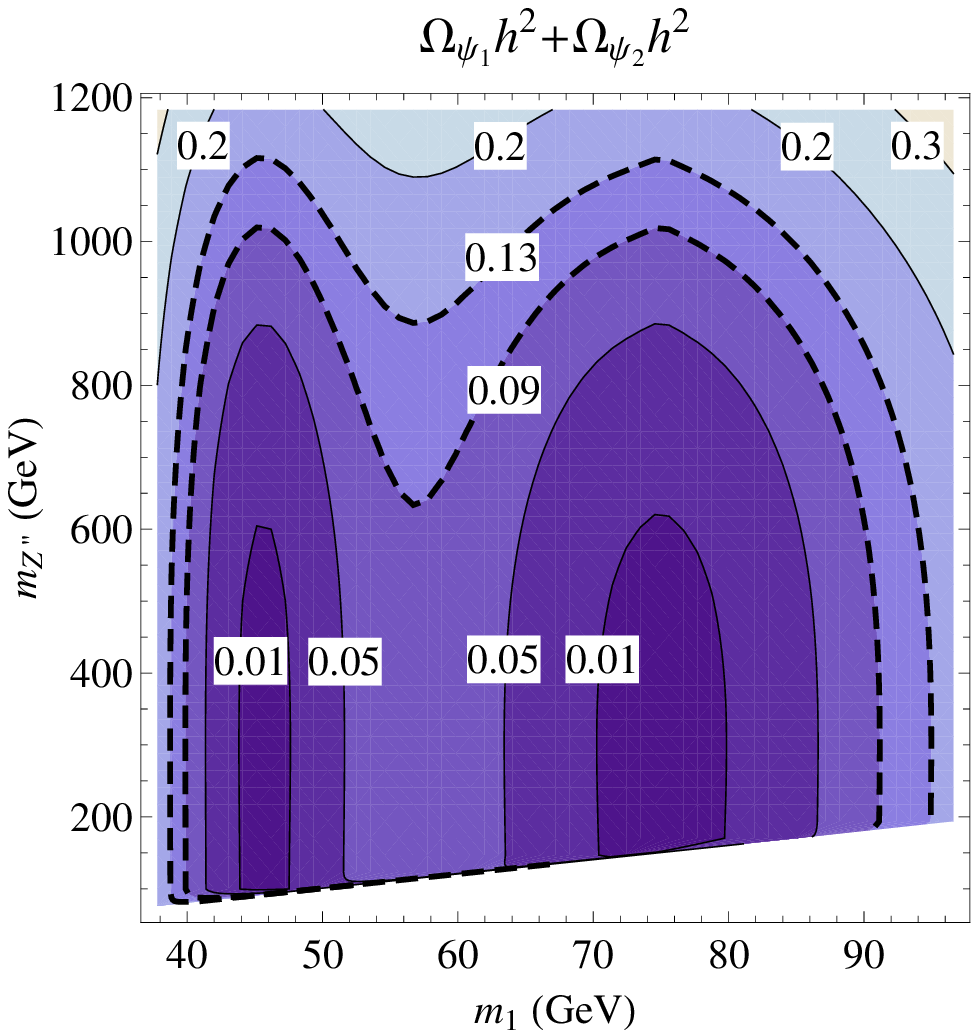}
\includegraphics[width=0.47\linewidth]{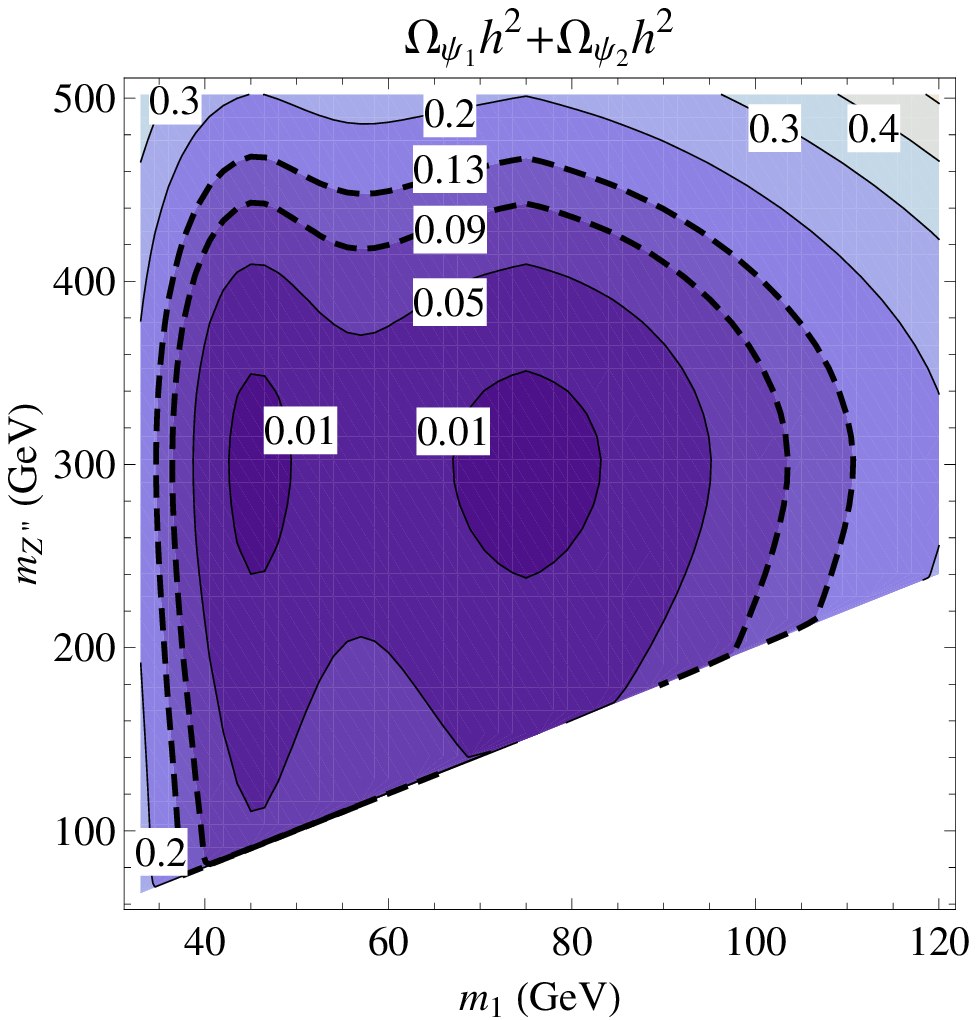}
\end{center}
\caption{Contour plots for the total relic abundance of the dark
matter particles $\psi_1$ and $\psi_2$ in the $m_1 - m_{Z''}$ plane.
The left and right panels correspond to $(g_{X'},g_{X''}) = (0.3,0.9)$
and $(0.5,0.3)$ respectively. The other parameters are fixed at the
reference parameter values. In each plane, the region between two
thick dashed lines is preferred by the WMAP DM relic density result.}\label{fig:gx12}
\end{figure}

The annihilations of $\psi_1$ and $\psi_2$ are controlled by the
hidden gauge couplings $g_{X'}$ and $g_{X''}$, respectively. To
study their effect, we reduce each coupling separately. In
fig.~\ref{fig:gx12}, the total DM thermal relic density
$\Omega_{\psi_1}h^2 + \Omega_{\psi_2}h^2$ is shown for the two
representative cases $g_{X'} = 0.3$ or $g_{X''} = 0.3$ in the $m_1
- m_{Z''}$ parameter space. For the other parameters, we take the
same values as in the reference case of fig.~\ref{fig:reference}.
Comparing the bottom panel of fig.~\ref{fig:reference} with
$(g_{X'}, g_{X''}) = (0.5, 0.9)$ with the left panel of
fig.~\ref{fig:gx12} with $(g_{X'}, g_{X''}) = (0.3, 0.9)$, one
sees that the allowed bands at low values of $m_{Z''}$ move closer
to either $m_Z/2\; {\rm or} \; m_{Z'}/2$. This is because as the
hidden gauge coupling $g_{X'}$ decreases, the annihilation of
$\psi_1$ is weaker, and stronger resonance effect is therefore
required to obtain appropriate annihilation strength. This also
means that in the mass region between the two resonances, $m_1
\approx 60$~GeV, the relic density of $\psi_1$ increases.
Therefore, the total DM density can be in agreement with the
measured value for lighter $Z''$ masses. When the hidden gauge
coupling is decreased to $g_{X''} = 0.3$, the requirement of a
stronger resonance effect means that the region in agreement with
the observed value of the relic density moves to lower values of
$m_{Z''}$, see the right panel of fig.~\ref{fig:gx12}.

%
%
%
%

\begin{figure}
\begin{center}
\includegraphics[width=0.47\linewidth]{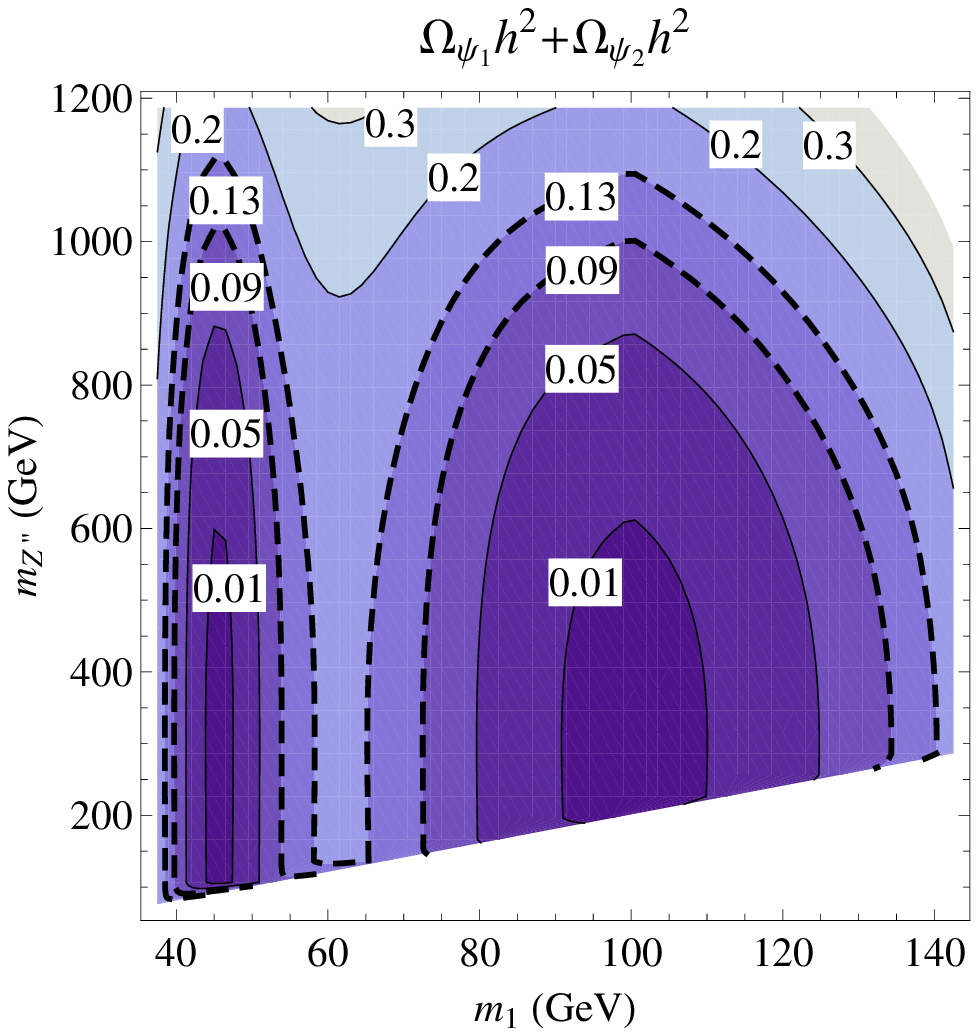}
\includegraphics[width=0.47\linewidth]{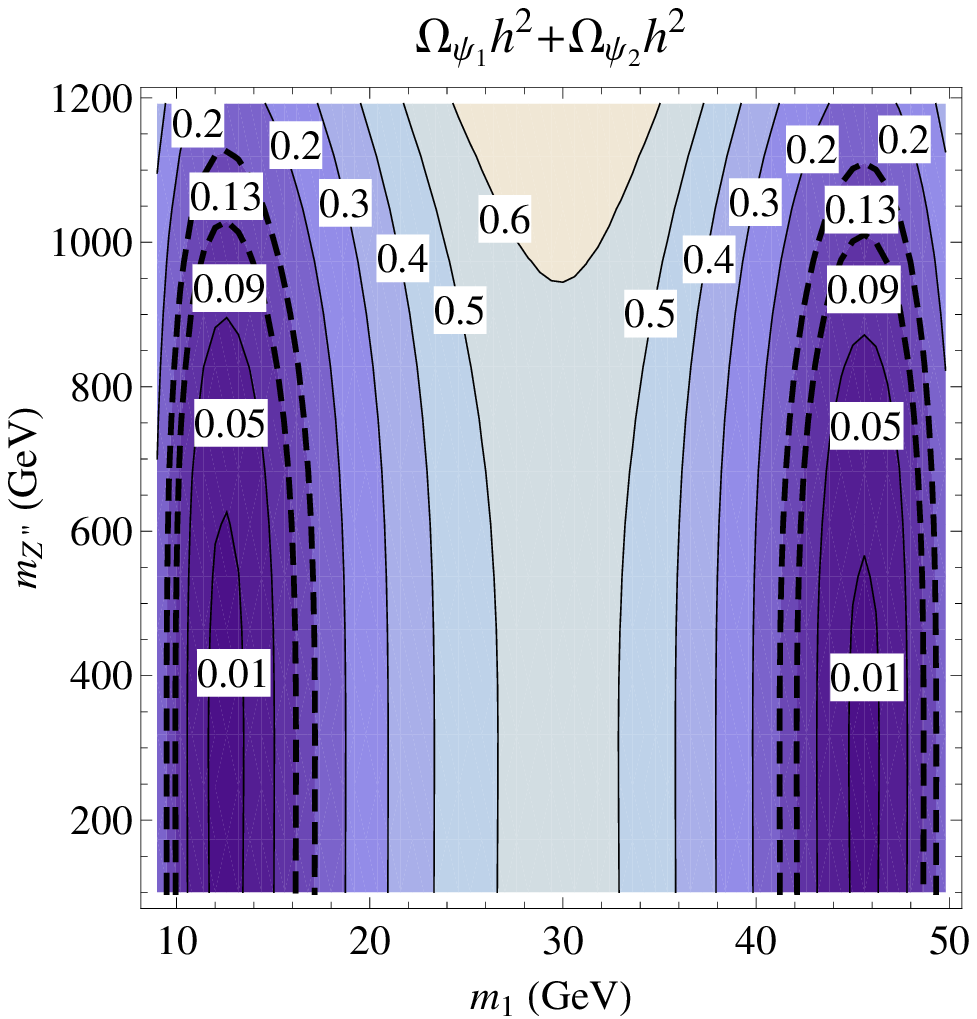}
\end{center}
\caption{Contour plots for the total relic abundance of the dark
matter particles $\psi_1$ and $\psi_2$ in the $m_1 - m_{Z''}$ plane
for the cases $m_{Z'} = 200$~GeV and $m_{Z'} = 25$~GeV. The parameters
$g_{X'}$, $g_{X''}$ and $m_2$ are fixed as the reference values. The
regions between two thick dashed lines are allowed by the recent DM
relic density observation.}\label{fig:mx2}
\end{figure}

The interactions of the DM particle $\psi_1$ also depend on the
$Z'$ mass $m_{Z'}$. In fig.~\ref{fig:mx2}, we display the total DM
relic abundance $\Omega_{\psi_1}h^2 + \Omega_{\psi_2}h^2$ in the
$m_1 - m_{Z''}$ parameter space to illustrate the dependence on
the $Z'$ mass $m_{Z'}$.
%
%
%
%
In the figure, the left panel corresponds to $m_{Z'} = 200$~GeV and the right panel to $m_{Z'}=25$ GeV. In this analysis, we used the reference values for the other parameters. As can be seen clearly from fig.~\ref{fig:mx2}, there are two well separated resonance regions, one around $m_Z/2$, the other
near  $m_{Z'}/2$. In addition, the right panel shows that the
lighter particle $\psi_1$ can be a DM candidate around 10 GeV when
$Z'$ is light. Thus, this model could have a light DM candidate as
hinted by some direct detection results, see more details in the
next section.

\begin{figure}
\begin{center}
\includegraphics[width=0.47\linewidth]{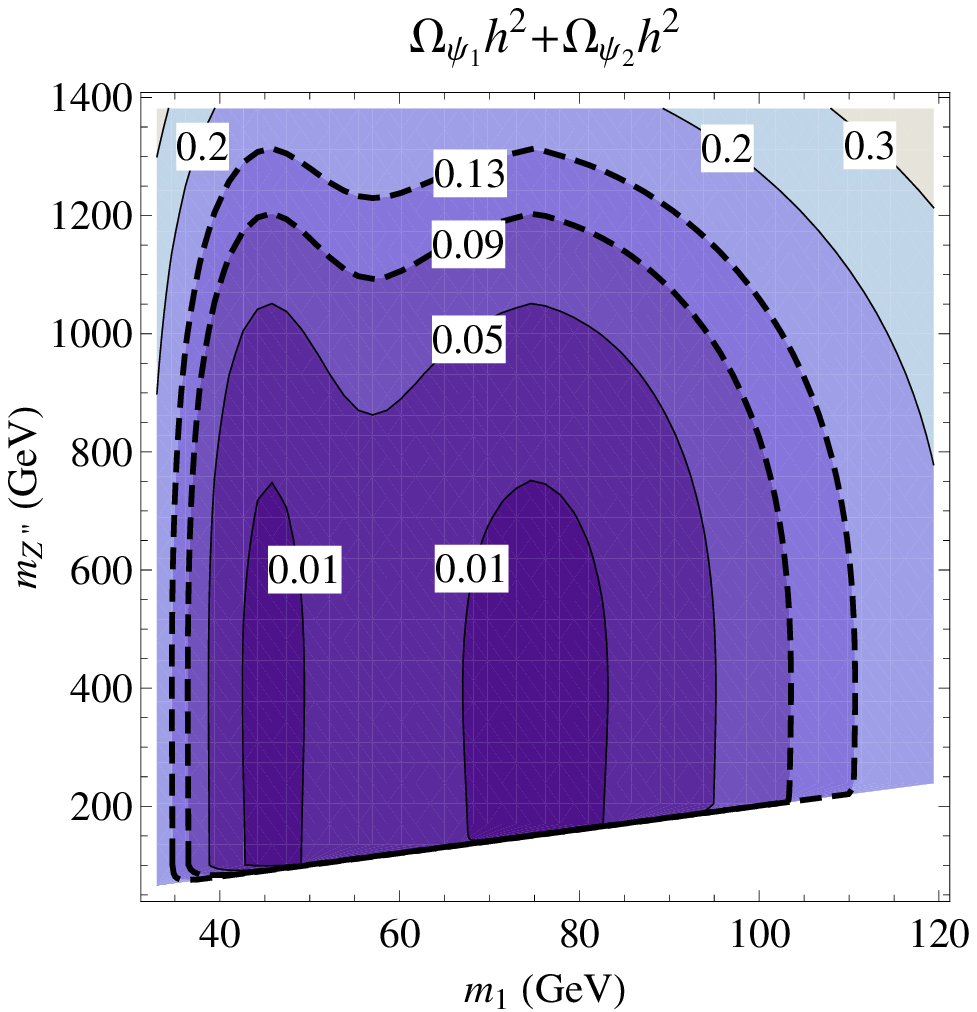}
\end{center}
\caption{Contour plots for the total relic abundance of the dark matter
particles $\psi_1$ and $\psi_2$ in the $m_1 - m_{Z''}$ plane for the case
$m_2 = 200$ GeV. The reference values are used for the other parameters.
The regions between two thick dashed lines are allowed by the recent DM
relic density observation.}\label{fig:m2}
\end{figure}

Finally, we investigate  the dependence of the DM relic density on
the $\psi_2$ mass $m_2$. In fig.~\ref{fig:m2}, the total DM relic
density of $\psi_1$ and $\psi_2$ is shown for the representative
case $m_2 = 200$ GeV in the $m_1 - m_{Z''}$ plane which is to be
compared with fig.~\ref{fig:reference} for which $m_2=150$~GeV.
All other parameters are kept to the reference values, $g_{X'} =
0.5$, $g_{X''} = 0.9$ and $m_{Z'} = 150$~GeV.
%
%
%
%
As expected, for larger $\psi_2$ mass, the preferred parameter
space moves to larger $m_{Z''}$ region, now $m_{Z''}\approx
1.2$~TeV due to the change of the $m_{Z''}$ resonance region while
the preferred value for $m_1$ at low values of $m_{Z''}$ is not
much shifted. One can see the tendency from the bottom panel of
fig.~\ref{fig:reference} and  fig.~\ref{fig:m2}.

\subsection{Direct detection}

We have shown that this simple model realizing the assisted
freeze-out mechanism can satisfy the observed DM relic abundance.
We now discuss the influence of this mechanism on the prospects of
observing DM in direct detection experiments. The most
distinguishing feature of the assisted freeze-out scenario
compared with the other multi-DM scenarios is that only lighter DM
particle $\psi_1$ can directly interact with the SM sector.
Consequently, only $\psi_1$ can be detected in DM direct detection
experiments. The lighter DM particle $\psi_1$ can elastically
scatter off a target nucleus through $t-$channel $Z$ and $Z'$
gauge boson exchange. One can easily calculate the
spin-independent (SI) $\psi_1$-nucleon cross section using the
following effective operator:
\begin{eqnarray}
&& {\cal L}_{eff} = b_f\, \bar{\psi_1}\gamma_\mu \psi_1\, \bar{f} \gamma^\mu f\,, \\
\mbox{where} &&
 b_f = {g^Z_{\psi_1} (g^Z_{fL}+g^Z_{fR}) \over 2 m_Z^2}
     + {g^{Z'}_{\psi_1} (g^{Z'}_{fL}+g^{Z'}_{fR}) \over 2 m_{Z'}^2} \,.
\label{eq:bf}
\end{eqnarray}

The current experimental bounds on $\sigma_n^{\rm SI}$ are
extracted from DM direct detection experiment results assuming
that the couplings to protons ($f_p$) and neutrons ($f_n$) are
equal. However, the couplings are different in this model. In
order to compare directly with the limits on $\sigma_n^{\rm SI}$
given by experiments, we thus use the normalized cross section on
a point-like nucleus~\cite{arXiv:1008.0580}:
\begin{eqnarray}\label{NormalizedCross}
\sigma_{\psi_1N}^{\rm SI} = \frac{\mu_{\psi_1}^2}{\pi}\, \frac{[Zf_p + (A-Z)f_n]^2}{A^2}\,.
\end{eqnarray}
Moreover, the experimental limits are also extracted assuming that
the local DM density is due to only one DM candidate. However, in
our model only the lighter particle $\psi_1$ can be observed by DM
direct detection experiments. Therefore, we rescale the SI
scattering cross section by $\Omega_{\psi_1}h^2 / \Omega_{\rm DM}
h^2$ assuming that the contribution of each DM particles to the
local density is the same as their contribution to the relic
density.

\begin{figure}
\begin{center}
\includegraphics[width=0.7\linewidth]{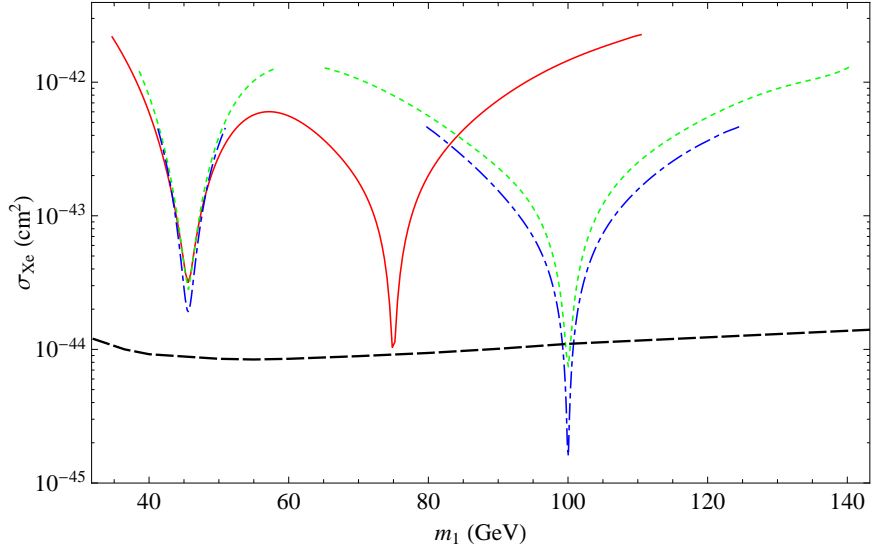}
\end{center}
\caption{SI scattering cross section of $\psi_1$ normalized to Xenon target nucleus
and rescaled by $\Omega_{\psi_1}h^2 / \Omega_{\rm DM} h^2$ (see text).
The red solid, green dotted and blue dot-dashed lines correspond to
$(g_{X'},~m_{Z'}) =$ (0.5, 150 GeV), (0.5, 200 GeV) and (0.3, 200 GeV),
respectively. $g_{X''}$ and $m_2$ are fixed to 0.9 and 150 GeV.
The experimental limit, which is taken from XENON100~\cite{XENON100},
is shown by the black dashed line. }
\label{fig:cross}
\end{figure}

In fig.~\ref{fig:cross}, we present the normalized SI scattering
cross sections of $\psi_1$ as a function of $m_1$ for
$(g_{X'},~m_{Z'}) =$ (0.5, 150 GeV), (0.5, 200 GeV) and (0.3, 200
GeV) fixing $g_{X''} = 0.9$ and $m_2 = 150$ GeV. The value of
$m_{Z''}$ is chosen to follow the contour $\Omega_{\psi_1}h^2 +
\Omega_{\psi_2}h^2 = 0.13$. The first feature is that the rescaled
cross section decreases around the resonance points ($m_1 \approx
m_Z/2,~m_{Z'}/2$), this is simply because $\Omega_{\psi_1}h^2$
decreases sharply near the resonances. Actually, the elastic cross
section itself has a much milder and smoother dependence on $m_1$.

%
%
%
%

Comparing the cases $(g_{X'}, m_{Z'}) =$ (0.5, 150 GeV) and (0.5,
200 GeV), one can see that the cross sections become smaller for
larger $m_{Z'}$ because of the suppressed contribution of the $Z'$
exchange diagram. Thus, one can recover a region where the elastic
cross section drops below the limit of XENON100~\cite{XENON100}.
Both the $Z$ and $Z'$ exchange diagrams are also suppressed for
smaller values of $g_{X'}$, we therefore obtain smaller cross
sections in this case as can be seen by comparing the curves for
$(g_{X'}, m_{Z'}) =$ (0.5, 200 GeV) and (0.3, 200 GeV). Note that
in fig.~\ref{fig:cross} there is a discontinuity in the direct
detection curves when $m_{Z'}=200$~GeV, which is because the
$\Omega h^2=0.13$ contour is also disconnected, see
fig.~\ref{fig:mx2}. In our analysis, we fix $\sin \epsilon$ to the
experimental upper limit as mentioned in the previous subsection.
If we use smaller $\sin \epsilon$, the effect is very similar to
the case of smaller $g_{X'}$ since the interactions between
$\psi_1$ and SM particles are approximately proportional to
$g_{X'} \sin \epsilon$. In summary, we therefore find that it is
easier to satisfy the direct detection constraint for smaller
values of $g_{X'} \sin \epsilon$ and larger values of $m_{Z'}$
than those of the reference case.

In this analysis, we apply the normalization of the cross section
to Xenon, Eq.~(\ref{NormalizedCross}), since XENON100 provides the
most stringent limit in most mass range. Actually, the cross
section for Germanium is very similar to the Xenon case due to
similar $Z/A$ ratios in both nuclei.

\begin{figure}
\begin{center}
\includegraphics[width=0.7\linewidth]{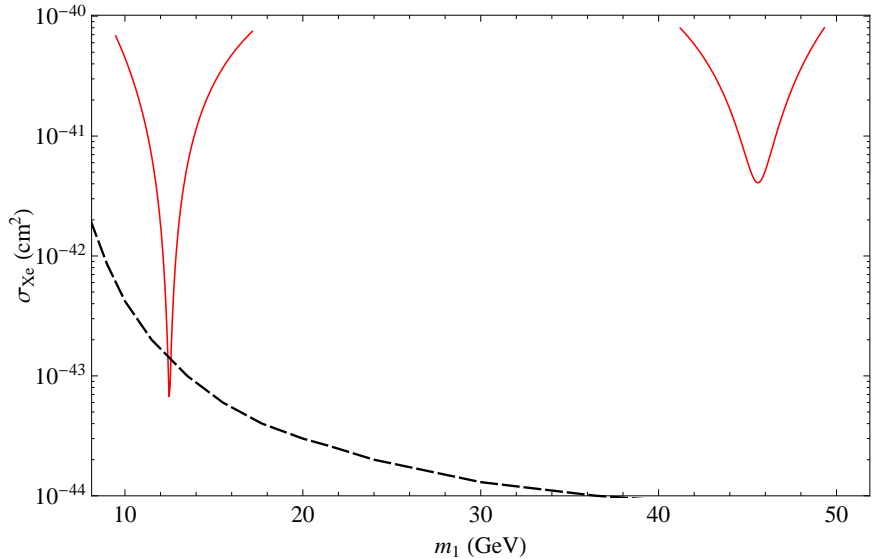}
\end{center}
\caption{SI scattering cross section of $\psi_1$ for the Xenon target
with $g_{X'} = 0.5$, $g_{X''} = 0.9$, $m_2 =$ 150 GeV and $m_{Z'} =$ 25 GeV.
The experimental limit from XENON100~\cite{XENON100} is shown by the
black dashed line. }\label{fig:mx25}
\end{figure}

In order to check the possibility of a light DM hinted by the data
from DAMA~\cite{DAMA}, CoGeNT~\cite{Cogent} and
CRESST~\cite{CRESST}, we show the SI scattering cross section
corresponding to a light $Z'$. We choose  $m_{Z'} =$ 25 GeV using
$g_{X'} = 0.5$, $g_{X''} = 0.9$ and $m_2 = 150$ GeV in
fig.~\ref{fig:mx25}.
%
%
%
%
As can be seen from the figure, the scattering cross section is
severely constrained by the XENON100 limit since the scattering
rate is enhanced due to the small mass of $Z'$. Nevertheless, we
can find small allowed region around the $Z'$ resonance point of
$m_1 \approx m_{Z'}/2$. One can easily expand this region, where
the SI cross section drops below the XENON100 bound, using smaller
$g_{X'} \sin \epsilon$. In addition, lighter DM particle $\psi_1$
can satisfy the scattering cross sections required by the DAMA,
CoGeNT or CRESST results.

\section{Conclusion}

We have illustrated with a simple toy model containing two stable
dark matter particles, how the assisted freeze-out mechanism
worked and could reproduce the measured value for the relic
density of dark matter. The main feature of this type of model is
that only one of the DM particles is involved in direct detection
searches while both contribute to the relic density. In
particular, when the DM particle that interacts with SM particles
is the subdominant DM component, it is possible to reconcile
models with large elastic scattering rates on nuclei with the
exclusion bounds of XENON100. Moreover, the lighter particle can
be a light DM candidate around 10 GeV as indicated  by the DAMA,
CoGeNT and CRESST results.

\acknowledgements{We thank the LPSC, Grenoble where part of this
work was carried out for their hospitality.}

\end{document}